\documentclass[a4paper,11pt]{article}
\usepackage{jheppub} 
\usepackage{lineno}
\usepackage{graphicx}
\usepackage{subcaption}
\usepackage{float}
\usepackage{tikz}
\usetikzlibrary{arrows}
\usetikzlibrary{decorations.markings}

\arxivnumber{} 

\title{\boldmath From Divergent Series to Geometry: Resurgence of the Quantum Metric}

\author[a]{Marcos J. Hern\'andez,}
\author[a,b]{Bogar D\'{\i}az,}
\author[a]{and J. David Vergara}

\affiliation[a]{Departamento de F\'isica de Altas Energ\'ias, Instituto de Ciencias Nucleares, Universidad Nacional Aut\'onoma de M\'exico, Apartado Postal 70-543, Ciudad de M\'exico, 04510, M\'exico}
\affiliation[b]{Group of Biometrics, Biosignals, Security and Smart Mobility (GB2S), E.T.S.I. Minas y Energía, Universidad Politécnica de Madrid, C. de Ríos Rosas, 21, 28003, Madrid, Spain}

\emailAdd{mhm@ciencias.unam.mx}
\emailAdd{bogar.diaz@upm.es}

\emailAdd{vergara@nucleares.unam.mx}

\abstract{ In this work, we analyze perturbative expansions of the quantum metric tensor (QMT) in anharmonic oscillators, focusing on quartic, sextic, and $d$-dimensional models. Using high-order perturbation theory, we show that the divergent QMT series exhibit factorial growth. Our analysis identifies universal non-perturbative scales, with coefficients displaying large-order behavior consistent with resurgence theory. Then, we apply resurgence and Borel--Padé resummation to the QMT. Comparisons with exact diagonalization confirm that Borel--Padé resummations yield accurate results, especially for the ground state. For completeness, we also present the analysis of the energy eigenvalues in the examples. Our findings extend resurgent techniques from energies to the QMT, highlighting the interplay between quantum geometry and non-perturbative physics.}

\begin{document}
\maketitle
\flushbottom

\section{Introduction}
\label{sec:intro}

Perturbation theory remains one of the most widely used methods in quantum mechanics and quantum field theory. However, in most physically relevant systems, the resulting perturbative expansions are not convergent but rather divergent asymptotic series~\cite{Dyson1952,BenderWu1969}. Despite their divergence, such expansions encode highly accurate information when truncated optimally, and—more importantly—their large-order behavior carries hidden connections with non-perturbative physics, such as instantons, tunneling amplitudes, or Stokes phenomena. Making this information explicit requires a resummation framework capable of transcending ordinary perturbation theory.

Resurgence theory provides precisely such a framework. Originating in the seminal works of Écalle~\cite{Ecalle1981}, it has been extensively developed in mathematical physics to uncover deep relations between perturbative and non-perturbative sectors~\cite{DelabaerePham1999,DunneUnsal2014,Marino2015,ANICETO}. At its core, resurgence rests on two key ideas: (i) divergent perturbative expansions are not pathological, but rather resurgent objects whose Borel transforms possess singularities encoding additional physical data; and (ii) suitable resummation procedures, such as Borel–Padé or generalized Borel–Leroy transforms, can reconstruct exact results from divergent series once singularity structures are properly accounted for. In this way, resurgence theory not only renders divergent series meaningful but also unveils the analytic and algebraic structures that govern them.

In parallel, the concept of quantum geometry has attracted increasing attention in recent years, motivated by both foundational aspects and applications ranging from quantum information to condensed matter physics~\cite{Provost,Zanardi2007Information,Carollo2020,Torma}. The central mathematical object is the quantum geometric tensor (QGT), whose real part defines a Riemannian metric (the QMT) on the parameter manifold and whose imaginary part corresponds to the Berry curvature~\cite{Berry1984}. The QGT provides a natural tool to study quantum phase transitions, fidelity susceptibilities, and dynamical responses. Nevertheless, perturbative expansions of the QGT often inherit the divergent character of the underlying wave functions and energies, raising the question of how to reliably extract non-perturbative information from their asymptotic series.

In this work, we bring together these two perspectives by applying resurgence and Borel resummation methods to the perturbative expansions of the QMT in anharmonic oscillators for the first time. We focus on three representative models: the quartic oscillator, the sextic oscillator, and the $d$-dimensional quartic oscillator with spherical symmetry. These systems are classical benchmarks for divergent perturbation theory~\cite{BenderWu1973,Simon1970,ZinnJustin1981}, while also serving as analytically tractable laboratories for testing resurgent methods. These methods have primarily been used to calculate the energy eigenvalues with very high precision \cite{Babenko1,BenderWu1973,ZinnJustin1981}. We remark that they have not been applied in the context of the QGT. Our analysis demonstrates that: a) For the quartic oscillator, the factorial divergence of both the energy spectrum and the QMT is controlled by a singularity at $\lambda = -k^{3/2}/3$, leading to instanton-like non-perturbative corrections. However, for positive $\lambda$, as in our study, that contribution does not appear on the physical amplitudes, even though in the negative case it becomes relevant \cite{BenderWu1973}, b) For the sextic oscillator, the large-order growth is governed by $\Gamma(2n+\beta)$ laws, requiring generalized Borel–Leroy transforms. We show that this framework successfully resums the divergent series and captures the underlying non-perturbative physics, and c) For the $d$-dimensional quartic oscillator, the dimensional dependence modifies the large-order structure but preserves Borel summability, demonstrating that resurgence techniques remain effective in multidimensional quantum systems. These results are compared with those obtained by exact diagonalization \cite{Okun, Gonzalez2024,Babenko1} of the models, which show that resurgence theory and resummation techniques yield accurate outcomes with fewer computational resources.

The work is organized as follows. In Section~\ref{sec:QMT}, we review the general structure of the QMT and introduce the notation used throughout the paper. Section~\ref{section:resurgence} summarizes the main ideas of resurgence theory and Borel resummation, including generalized Borel transforms. In Sections~\ref{sec:quartic}, \ref{sec:sextic}, and \ref{sec:ddim}, we apply these methods to the quartic oscillator, the sextic oscillator, and the $d$-dimensional quartic oscillator, respectively. We conclude in Section~\ref{sec:conclusions} with a discussion of the implications of our results and possible extensions to more general quantum systems.

\section{Quantum Metric Tensor}\label{sec:QMT}

In this section, we present some fundamental concepts concerning the geometry of parameter spaces for quantum systems and establish the notation used throughout this work.

Let us consider a quantum system in one spatial dimension, governed by a Hamiltonian $\hat{H}(\hat{q}, \hat{p}; \lambda)$ that depends smoothly on a set of $\mathcal{N}$ real parameters $\lambda = \{\lambda^i\}$, with $i = 1, \ldots, \mathcal{N}$. We assume that, for each value of the parameters, the Hamiltonian admits a non-degenerate eigenstate $|\Psi_N (\lambda)\rangle$ with eigenvalue $E_N (\lambda)$.

The geometry of the parameter space $\mathcal{M}$ can be characterized by introducing the QMT, which quantifies infinitesimal distances between neighboring eigenstates. Following the approach of Provost and Vallee~\cite{Provost}, the components of the QMT are given by
\begin{equation}
g_{ij}^{\scriptscriptstyle (N)} := \mathrm{Re} \left( \langle \partial_i \Psi_N | \partial_j \Psi_N \rangle - \langle \partial_i \Psi_N | \Psi_N \rangle \langle \Psi_N | \partial_j \Psi_N \rangle \right)\,, \label{QMT}
\end{equation}
where $\partial_i \equiv \frac{\partial}{\partial \lambda^i}$. This Riemannian metric defines the squared line element $\delta \ell^2 = g_{ij}^{\scriptscriptstyle (N)}(\lambda) \delta
\lambda^i \delta \lambda^j$ in parameter space, thus capturing how the eigenstate $|\Psi_N(\lambda)\rangle$ varies under small changes in the parameters $\lambda^i$.

A complementary and often more computationally accessible representation of the QMT arises from first-order perturbation theory~\cite{Zanardi2007Information}. It reads as follows
\begin{equation}
g_{ij}^{\scriptscriptstyle (N)}= \mathrm{Re} \sum_{M \ne N} \frac{\langle \Psi_N | \hat{O}_i | \Psi_M \rangle \langle \Psi_M | \hat{O}_j | \Psi_N \rangle}{(E_M - E_N)^2}\,, \label{QMTpert}
\end{equation}
where we define $\hat{O}_i := \partial_i \hat{H}$. This expression highlights the sensitivity of the metric tensor to level crossings, since it becomes divergent when $E_M(\lambda) = E_N(\lambda)$, indicating the presence of critical points, such as those encountered in quantum phase transitions. Nevertheless, a thorough analysis is often required to determine whether such singularities are physically meaningful~\cite{Zanardi2007Scaling, Carollo2020}. 

Beyond its geometric definition, the QMT plays a central role in various domains of quantum theory.  The QMT quantifies the distinguishability of nearby quantum states and determines fidelity susceptibility~\cite{Gu2010}, which is a key diagnostic of quantum phase transitions. In addition, it appears in the theory of adiabatic response and quantum thermodynamics, where it governs energy fluctuations and work statistics~\cite{Kolodrubetz2017}. More recently, it has been explored in contexts ranging from quantum information geometry \cite{Carollo2020}, quantum field theory \cite{Johanna}, and condensed matter physics \cite{Yu} highlighting its significance as a probe of both geometric and dynamical properties of quantum systems.

\section{General idea of resurgence in quantum mechanics} \label{section:resurgence}

The resurgence theory establishes a profound connection between perturbative and non-perturbative phenomena in quantum theories. It is used to reveal a hidden algebraic structure that bridges asymptotic series expansions with exact non-perturbative solutions. 

Consider a formal power series of the form
\begin{align}
    \phi(z)=\sum_{n=0}^{\infty}a_n z^n\,. \label{eq:divser}
\end{align}
We said that it is an asymptotic approximation to the function $f(z)$ in the sense of Poincaré, denoted by $f(z) \sim \phi(z) $, if for all $A>0$
\begin{align}
    \lim_{z\to 0} z^{-A} \left( f(z) -\sum_{n=0}^{A}a_n z^n\right)=0\,.
\end{align}

Notice that we are not demanding $\lim_{A\to\infty} \left( f(z) -\sum_{n=0}^{A}a_n z^n\right)=0$ for fixed $z$. This behavior differs from that of a convergent series, where both limits must approach zero. Even when the series \eqref{eq:divser} may diverge, this asymptotic approximation could provide excellent estimates when truncated at an optimal order, at least for certain coupling constant values, but typically fails for others.  The optimal order approach uses only a finite number of terms from the asymptotic approximations, leaving the remaining terms unexploited for improving the approximation (except for convergent series). A better approach to incorporating the information from all terms in the asymptotic approximations is accomplished through \emph{Borel resummation} (resurgence theory). 

We are interested in perturbative asymptotic series that satisfy the Gevrey-1 condition, i.e.,  their coefficients satisfy
\begin{equation}
|a_n| \leq C \, R^n \, n! \quad \text{for some } C, R > 0\,.
\end{equation}

For Borel resummation, we first require the Borel transform and an analytic continuation. Notice that perturbation theory in quantum mechanics typically produces formal power series in a coupling or $\hbar$ that are \emph{divergent} but asymptotic. The basic steps of resurgence theory are \cite{Jentschura,Marinonet,ANICETO}:

\begin{enumerate}

\item  \textbf{Borel transform}: given a factorially divergent asymptotic approximation $f(z) \sim \sum_{n=0}^\infty a_n z^n$ with $a_n \sim n!$, the Borel transform of $f$ is defined as:

\begin{equation}
\mathcal{B}[f](u) := \sum_{n=0}^\infty \frac{a_n}{n!} u^n \,. \label{Boreltransform}
\end{equation}

The $\mathcal{B}[f](u)$ typically converges in a finite disk around $u=0$, but it can have singularities (and then poles). 
One of the key ideas of resurgence is that these singularities can contain information about additional sectors of the theory. Then, the Borel transform has to be continued analytically beyond the radius of convergence. However, such a continuation requires the knowledge of all coefficients of the series, which is rather uncommon in physics. Fortunately, if we know a finite number of coefficients of the series, an efficient way to produce an approximate analytic continuation is the \emph{Pad\'e approximant}. We must mention that other approaches to the analytic continuation exist (see, for example \cite{Mera2018}).

\item \textbf{Pad\'e Approximation of the Borel Transform}:  In general, the $[P/Q]$ Pad\'e approximant \cite{baker1996pade} of a series $\phi$ is the unique rational function:
\begin{equation}
\left[\frac{P}{Q} \right]_\phi (u) = \frac{\sum_{m=0}^P p_m u^m}{1 + \sum_{n=1}^Q q_n u^n} \,,
\end{equation}
satisfying the matching condition:
\begin{equation}
\left[\frac{P}{Q} \right]_\phi (u) - \phi(u) = \mathcal{O}(u^{P+Q+1}) \,.
\end{equation}
Then, the Padé approximant produces a rational approximation to the original input series given by the ratio of two polynomials. If we have the first $m+1$ terms of the Borel transform  \eqref{Boreltransform} (as it is usually in physics), we can construct a diagonal or off–diagonal Padé approximant  of the Borel transform $\mathcal{B}[f](u)$, which is given by

\begin{equation}
\mathcal{P}_{m}[f](u) := \left[ \frac{[[\frac{m}{2}]]} {[[ \frac{m+1}{2}]]}\right]_{\mathcal{B}[f]} (u)\,, \label{padelapprox}
\end{equation}
where $[[\cdot]]$ denotes the integer part. Notice that we can also compute the Padé approximant of the function, which we denote by
\begin{equation}
P_{m}[f](u) := \left[ \frac{[[\frac{m}{2}]]} {[[ \frac{m+1}{2}]]}\right]_{f} (u)\,. \label{padelapprox}
\end{equation}
Usually, this approximant improves the convergence of the series, and its value is closer to the exact result.

\item \textbf{Modified Laplace transformation}: After an analytic continuation of $\mathcal{B}[f]$ we define the (ordinary) Borel sum
\begin{equation}
\mathcal{TP}_m [f] (z)= \frac{1}{z} \int_{0}^{\infty} \mathcal{P}_m [f] (u) \exp(-u/z) \, \mathrm{d}u\,,
\label{eq:BorelSum}
\end{equation}
which, when it exists, is analytic in a sector of the $z$‑plane and reproduces $f(z)$ asymptotically. When the sum converges, we say that the series is Borel summable.

In many physical problems $\mathcal{P}_m [f]$ has singularities on the positive real axis, so the integral \eqref{eq:BorelSum} is ill‑defined. However, we can select a contour that avoids the obstruction by an infinitesimal rotation, leading to the \emph{lateral} Borel sums

\begin{align}
        \mathcal{T P}_m^{\pm} [f] (z)=\frac{1}{z} \int_{C_{\pm}} \mathcal{P}_m  [f](u) \exp(-u/z) \, \mathrm{d}u\,.\label{eq:lateral}
\end{align}

In Fig. \ref{fig:BorelContoursThree}, we show the contours involved in \eqref{eq:lateral}.
The arithmetic mean of the results of the integrations along $C_-$ and $C_+$ is associated with $C_0$.

\begin{figure}[h]
\centering
\begin{tikzpicture}[>=stealth, thick, scale=1.0]
  \draw[->] (-0.5, 0) -- (8, 0) node[below right] {$\textrm{ Re}\,u$};
  \draw[->] (0, -2.5) -- (0, 2.5) node[left] {$\textrm{Im}\,u$};

\draw[line width=2pt, red!80] (0, 0) -- (7.5, 0) node[above=2pt, midway, red]{};


  \foreach \x/\label in {2/A_1, 4.5/A_2, 7/A_3} {
  \filldraw[red] (\x, 0) circle (2.2pt);
}

 \filldraw[red] (1.5, 1) circle (2.2pt) node[above right] {};
 \draw[green!60!black, very thick] (1.5, 1) circle (0.25cm); 
 \filldraw[red] (5.5, 0.8) circle (2.2pt) node[above right] {};
 \draw[green!60!black, very thick] (5.5, 0.8) circle (0.25cm); 

 \filldraw[red] (3, -1.2) circle (2.2pt) node[below right] {};
 \draw[magenta, dashed, very thick] (3, -1.2) circle (0.25cm); 
 \filldraw[red] (6.5, -0.7) circle (2.2pt) node[below right] {};
 \draw[magenta, dashed, very thick] (6.5, -0.7) circle (0.25cm); 

  \draw[green!60!black, very thick, ->]
    (0.3, 0.25) -- (1.85, 0.25)
    arc (180:0:0.15) -- (4.35, 0.25)
    arc (180:0:0.15) -- (6.85, 0.25)
    arc (180:0:0.15) -- (7.6, 0.25)
    node[above left] {$C_{+1}$};

  \draw[magenta, dashed, very thick, ->]
    (0.3, -0.25) -- (1.85, -0.25)
    arc (-180:0:0.15) -- (4.35, -0.25)
    arc (-180:0:0.15) -- (6.85, -0.25)
    arc (-180:0:0.15) -- (7.6, -0.25)
    node[below left] {$C_{-1}$};

  \draw[blue, dotted, thick]
    (0.3, 0.08) -- (1.85, 0.08)
    arc (90:-90:0.08) -- (0.3, -0.08);
  \draw[blue, dotted, thick]
    (2.15, 0.08) -- (4.35, 0.08)
    arc (90:-90:0.08) -- (2.15, -0.08);
  \draw[blue, dotted, thick]
    (4.65, 0.08) -- (6.85, 0.08)
    arc (90:-90:0.08) -- (4.65, -0.08);
  \draw[blue, dotted, thick]
    (7.15, 0.08) -- (7.6, 0.08)
    node[below right, blue] {$C_0$};

  \begin{scope}[shift={(5.2,2.3)}]
    \draw[green!60!black, thick] (0,0) -- ++(0.5,0) node[right] {$C_{+}$ (Upper side)};
    \draw[magenta, dashed, thick] (0,-0.4) -- ++(0.5,0) node[right] {$C_{-}$ ( bottom side)};
    \draw[blue, dotted, thick] (0,-0.8) -- ++(0.5,0) node[right] {$C_0$};
  \end{scope}
\end{tikzpicture}
\caption{Borel resummation contours. The red points represent the possible singularities of the Padé approximant. The lateral contours \(C_{\pm}\) avoid the branch cut (red line) from above and below, yielding the lateral sums \(\mathcal{T}^{\pm}\), respectively.}
\label{fig:BorelContoursThree}
\end{figure}
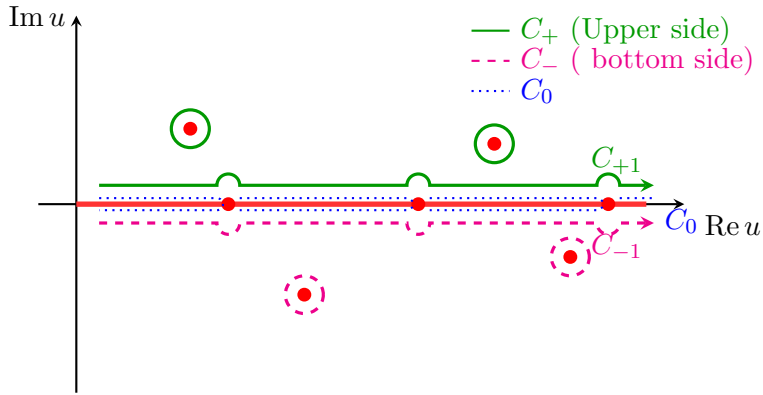

The $+$ ($-$) prescription integrates just above (below) the cut; the two can differ by an exponentially small discontinuity (per each pole). The difference can be a sum of different series with an exponentially small pre-factor. This indicates that to recover the function $f$, one must promote the series $\phi (z)$ to a trans-series (semiclassical decoding) \cite{Jentschura, Marinonet}. In physical theories, these contributions are the so-called instanton contributions.

The simplest situation corresponds to the case in which there are no singularities along the positive real axis, and the (ordinary) Borel resummation of the perturbative series reproduces the function $f$.
\end{enumerate}

\subsection{Generalised Borel transforms and non-standard factorial growth}

In the beginning of Section \ref{section:resurgence}, we considered the usual Gevrey-1 behaviour \(a_n\sim C\, R^n n!\).  In practice, however, perturbative coefficients may exhibit different large-order laws. For instance, we can encounter growth of the form \((k n)!\), with integer \(k>1\), factorials with an offset, or more generally
\begin{equation}
a_n \;\sim\; C\,R^n\,\Gamma(\alpha n+\beta)\,,\qquad \alpha>0,\ \beta\in\mathbb{R},
\label{eq:general_growth}
\end{equation}
so that the naive Borel transform \eqref{Boreltransform} is not the natural object. 

\paragraph{Generalised (\/\(\alpha\)-\/) Borel transform.} When the growth is of the form \(\Gamma(\alpha n+\beta)\) with \(\alpha\neq1\), one may define the generalized Borel transform \cite{Caliceti}
\begin{equation}
\mathcal{B}^{(\alpha,\beta)}[\varphi](u):=\sum_{n=0}^{\infty}\frac{a_n}{\Gamma(\alpha n+\beta)}\,u^n\,.
\label{eq:alphaBorel}
\end{equation}
A convenient inverse representation (valid under the usual analyticity and growth hypotheses \cite{Caliceti}) is 
\begin{equation}
\mathcal{T} \mathcal{B}^{(\alpha,\beta)}[\varphi](z)
=\frac{1}{\alpha z}\int_0^{\infty} \mathcal{B}^{(\alpha,\beta)}[\varphi](u) \,
\exp \left(-\left(\frac{u}{z}\right)^{1/\alpha}\right) \left( \frac{u}{z}\right)^{\beta/\alpha-1}\,
\,\mathrm{d}u\,, \label{eq:alphaInverse}
\end{equation}
which reduces to the ordinary Borel formulas of the Section \ref{section:resurgence} when \(\alpha =1= \beta\). Once again, if singularities of \(\mathcal{B}^{(\alpha,\beta)}\) lie on the positive real \(s\)-axis, one must use lateral deformations.

\paragraph{Practical implementation.} In numerical and semi-analytic work, one typically (i) divides the coefficients by the appropriate reference factor (e.g. \(\Gamma(\alpha n+\beta)\)), (ii) to effect analytic continuation, one constructs a Pad\'e approximant of the truncated transformed series (the notation for the Padé approximant after applying the generalized Borel transform \eqref{eq:alphaBorel} to a function $f$ is denoted by $\mathcal{P}^{(\alpha,\beta)}_{m}\left[f\right](u)$), and (iii) evaluates the corresponding Laplace-type integral with the lateral prescription if necessary.  

In the next section, we apply the resurgence theory to the perturbative expansion of the QMT in some examples. There, we will encounter several of these situations.  Whenever the large-order analysis indicates non-standard factorial growth, we will explicitly state the transform used (Borel or \(\alpha\)–Borel), apply Padé continuation to the transformed series, and perform the appropriate Laplace-type inversion (including lateral summation where required).  The differences among these choices will be exhibited in detail in the examples.

\section{Quantum quartic oscillator}\label{sec:quartic}

Let us consider, as our first example, the quartic potential. The corresponding quantum Hamiltonian is given by
\begin{align}
\label{cuartico}
    \hat{H}=\frac{\hat{p}^2}{2}+\frac{k \hat{q}^2}{2}+\lambda\hat{q}^4\,,
\end{align}
where $k$ and $\lambda$ are the adiabatic parameters of the system, i.e., $\lambda^i=(k,\lambda)$ in the general theory of Section \ref{sec:QMT}. This system has been extensively studied in the context of perturbation theory and divergent series. From the work of Bender and Wu \cite{BenderWu1969, BenderWu1973}, it is known that the perturbative expansion of energy levels in powers of $\lambda$ is a non-convergent asymptotic series, whose high-order behavior grows factorially. Then it is a candidate to apply the ideas of resurgence.

Actually, despite this divergence, the energy spectrum can be accurately obtained by Borel resummation, as it was shown in \cite{Simon1970, ZinnJustin1981}. For completeness, we present that analysis below. We must mention that in this case, the use of advanced techniques such as trans-series or resurgence theory is not strictly necessary. However, they provide a more comprehensive description of the analytical structure of the problem and its connection to non-perturbative effects \cite{DelabaerePham1999,DunneUnsal2014,Marino2015}.

The energy of the $N$-th excited state for the system \eqref{cuartico} can be expanded as a perturbative series in powers of $\lambda$ as

\begin{align}
    E^{(N)} = \left(N+\frac{1}{2}\right)\sqrt{k} + \sqrt{k} \sum_{n=1}^{\infty} a_{n}^{(N)} \left(\frac{\lambda}{k^{3/2}}\right)^n (-1)^{n+1}\,, \label{energycuartic}
\end{align}
where the coefficients $a_{n}^{(N)}$ are determined through Rayleigh--Schrödinger perturbation theory. Such perturbative expansions have been extensively investigated since the seminal works of Bender and Wu~\cite{BenderWu1969, BenderWu1973}, where they also analyzed the perturbative structure of cubic and quartic anharmonic oscillators and more recently by Babenko, et al \cite{Babenko2, Babenko1}

For the ground state ($N=0$), the first few coefficients are explicitly given by

\begin{align}
    a_{0}^{(0)} = \frac{1}{2}\,,\quad 
    a^{(0)}_{1} = \frac{3}{4}\,, \quad 
    a^{(0)}_{2} = \frac{21}{8}\,, \quad 
    a^{(0)}_{3} = \frac{333}{16}\,, \quad 
    a^{(0)}_{4} = \frac{30885}{128}\,, \quad 
    a^{(0)}_{5} = \frac{916731}{256}\,.
\end{align}

Bender and Wu also showed that the perturbative coefficients grow factorially as

\begin{align}
    a^{(0)}_n \sim \frac{\sqrt{6}}{\pi^{3/2}}\, 3^n\,  \Gamma\!\left(n+\tfrac{1}{2}\right)\,, \label{eq:coefenergy}
\end{align}
which implies the existence of a singularity at $\frac{\lambda}{k^{3/2}} = -\frac{1}{3}$ that determines the radius of convergence of the perturbative expansion. This asymptotic behavior is also reflected in the poles of the Padé approximants constructed from the (truncated at order $\lambda^m$) perturbative series. As the order of the approximant increases, the poles organize themselves into a regular pattern, forming a well-defined structure in the complex plane. Such behavior is characteristic of divergent expansions in quantum theories and has been extensively documented in analogous systems.

For the practical computation of these coefficients, we have made use of the \texttt{BenderWu} package~\cite{Sulejmanpasic2018BenderWu}, which efficiently implements the perturbative algorithm to very high orders. In particular, the ground-state energy has been calculated up to order 200 in $\lambda$, allowing for a detailed analysis of the analytic structure of the series.

The factorial growth of the perturbative coefficients and the associated singularity structure are precisely the ingredients that make this model a paradigmatic example of resurgence. In fact, the large-order behavior encodes non-perturbative information, which can be systematically extracted through Borel resummation techniques. In this framework, the singularity controls the emergence of exponentially small contributions of the form $\exp(-A/\lambda)$, linking the perturbative expansion with instanton-like non-perturbative effects (see below). Thus, the quartic oscillator provides an explicit and tractable realization of the general ideas discussed in the Section \ref{section:resurgence} on resurgence theory.

As discussed above, the perturbative expansion is divergent but asymptotic, making it natural to apply Borel resummation techniques to extract physically meaningful values. To this end, we consider the energy \eqref{energycuartic} but truncated at order $\lambda^m$ and we denote it by $E^{(0)}_{m}$. Then, we first apply to it the Borel transform \eqref{Boreltransform}, then, after constructing a Padé approximant $\mathcal{P}_m \left[ E^{(0)} \right]$ of the truncated Borel transform, the resummed energy $\mathcal{TP}_m[E^{(0)}]$ is obtained via the Laplace integral (ordinary Borel sum) \eqref{eq:BorelSum}.  Notice that we have omitted the subscript $m$ on the energy for the Padé approximant and the Borel sum; this is because the Padé notation already contains the information of the order of the considered series.

\begin{figure}[H]
\begin{subfigure}{0.45\linewidth}
    \centering
    \includegraphics[width=0.9\linewidth]{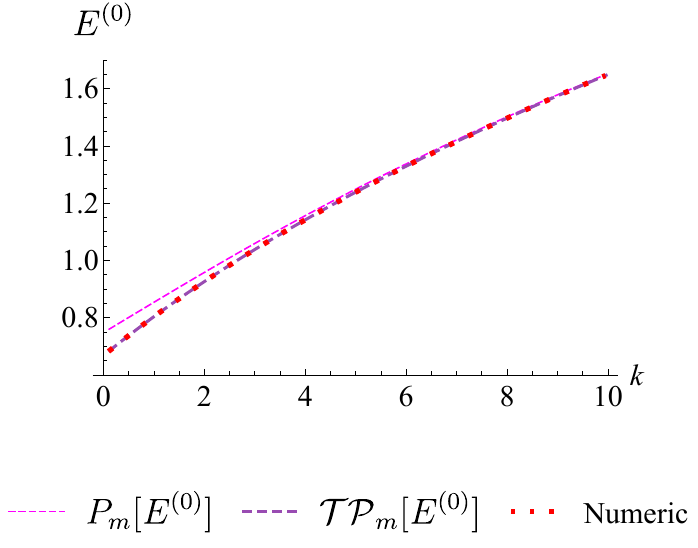}
    \caption{}
     \label{resurenerplota}
\end{subfigure}
\begin{subfigure}{0.45\linewidth}
    \centering
    \includegraphics[width=0.9\linewidth]{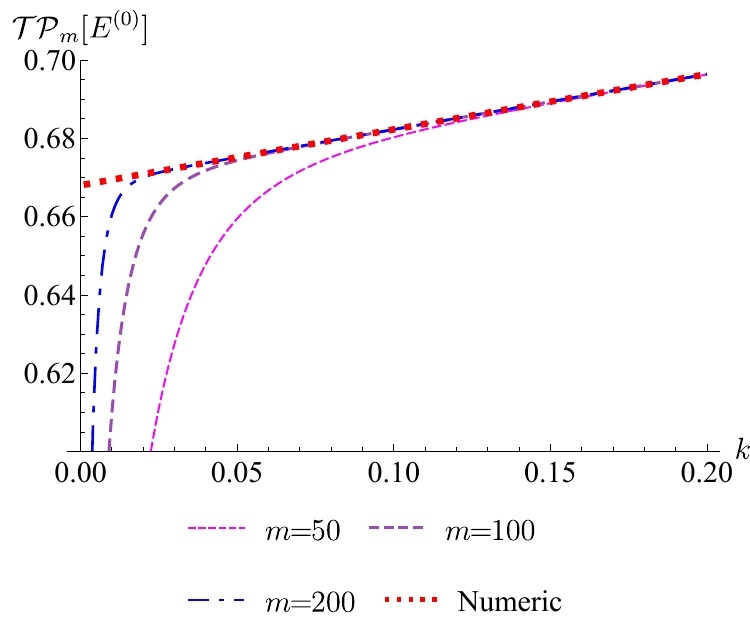}
    \caption{}  
    \label{resurenerplotb}
\end{subfigure}
\caption{(a) Ground-state energy computed using the Padé approximant, $P_m[E^{(0)}]$, and the Borel--Padé resummation $\mathcal{TP}_m[E^{(0)}]$ (with $m=100$), compared with the exact numerical energy obtained from Hamiltonian diagonalization. (b) Ground-state energy obtained from the Borel--Padé resummation $\mathcal{TP}_m[E^{0}]$ for $m=50,100,200$, compared with the exact diagonalization result.}
\label{resurenerplot}
\end{figure}

Next, we compare the energy obtained from the Padé approximant and Borel resummation with the exact energy computed by direct numerical diagonalization of the Hamiltonian \eqref{cuartico}. For the Padé and Borel–Padé resummations, we used $m=100$, while the numerical diagonalization was performed using a basis of size $s=200$. The results are presented in Fig.~\ref{resurenerplota}, where we fix $\lambda=1$ and vary $k$ in the interval $(0.1,10)$. It can be observed that the Padé approximant deviates from the exact energy as $k \to 0$, while the Borel resummation provides a more accurate approximation. To further illustrate this behavior, we show in Fig.~\ref{resurenerplotb}   the Borel resummed energy $\mathcal{TP}_m[E^{(0)}]$ for $m=50,100,200$, compared with the exact energy obtained from diagonalization. We observe that, as $m$ increases, the Borel resummation systematically improves and converges towards the exact result.

We now proceed to the perturbative calculation of the QMT. The wave function of the state $N$ was computed using the \texttt{BenderWu} Mathematica package, and, to obtain the QMT, we apply to it the Provost formula \eqref{QMTpert}. We obtain the components of the metric in the form
\begin{subequations} 
\label{metricacuartico}
\begin{align}
    g_{11}^{\scriptscriptstyle (N)}&=\frac{1}{k^2}\sum_{n=0}^{\infty}c^{\scriptscriptstyle (N;n)}_{11}\left( -\frac{\lambda}{k^{3/2}}\right)^{n} \, ,\\
    g_{12}^{\scriptscriptstyle (N)}&=\frac{1}{k^{5/2}}\sum_{n=0}^{\infty}c^{\scriptscriptstyle (N;n)}_{12}\left( -\frac{\lambda}{k^{3/2}}\right)^{n} \,,\\
    g_{22}^{\scriptscriptstyle (N)}&=\frac{1}{k^{3}}\sum_{n=0}^{\infty}c^{{(\scriptscriptstyle N;n)}}_{22}\left(- \frac{\lambda}{k^{3/2}}\right)^{n}\,.
\end{align}
\end{subequations}
For the ground state, the calculations were performed up to order 100. For the excited states, $N=1,2,3,4$, they were carried out up to order 50.  For the ground state, we find that the perturbative coefficients of the metric \eqref{metricacuartico} exhibit large-order behavior of the form 
\begin{subequations}
\label{coef:cuartico}
    \begin{align}
    c_{11}^{(n)}& \sim \frac{ 3^{n +\tfrac{5}{2}} }{4 \sqrt{2} \pi ^{3/2}} \Gamma \!\left(n +\tfrac{5}{2}\right)\,,\\
    c_{12}^{(n)}& \sim\frac{5  \cdot 3^{n+\tfrac{3}{2}} }{4 \sqrt{2} \pi ^{3/2}}\Gamma \!\left(n+\tfrac{7}{2}\right)\,,\\
    c_{22}^{(n)}& \sim\frac{19 \cdot 3^{n+\tfrac{3}{2}} }{10 \sqrt{2} \pi ^{3/2}}\Gamma \!\left(n+\tfrac{9}{2}\right)\,.
\end{align}
\end{subequations}
Thus, the perturbative QMT coefficients follow the generic large-order pattern 
\begin{align}
 c_{ij}^{(n)} \sim (-1)^n S_{ij}\,A^{-n}\Gamma(n+\beta_{ij})\,, \qquad A=1/3\,, \label{eq:coeffQMT}
\end{align}
with $\beta_{11}=5/2$, $\beta_{12}=7/2$, $\beta_{22}=9/2$, and $S_{ij}$ a numeric factor depending on the component. By standard resurgence arguments \cite{LeGuillou1990,Dorigoni}, this implies non-perturbative corrections of the form
\[
g_{ij}^{{\scriptscriptstyle (\mathrm{NP})}}\sim C_{ij}\,\lambda^{-\beta_{ij}}\,e^{1/(3\lambda)}\,,
\]
for the ground state. This is consistent with the accumulation of Padé–Borel poles at $u=-1/3$.

In the following subsections, we apply the resurgence ideas to these metric components for the ground state and some excited states.

\subsection{Borel resummation for the metric components  for the ground state}

\begin{figure}[H]
\centering
{\includegraphics[width=0.8\linewidth]{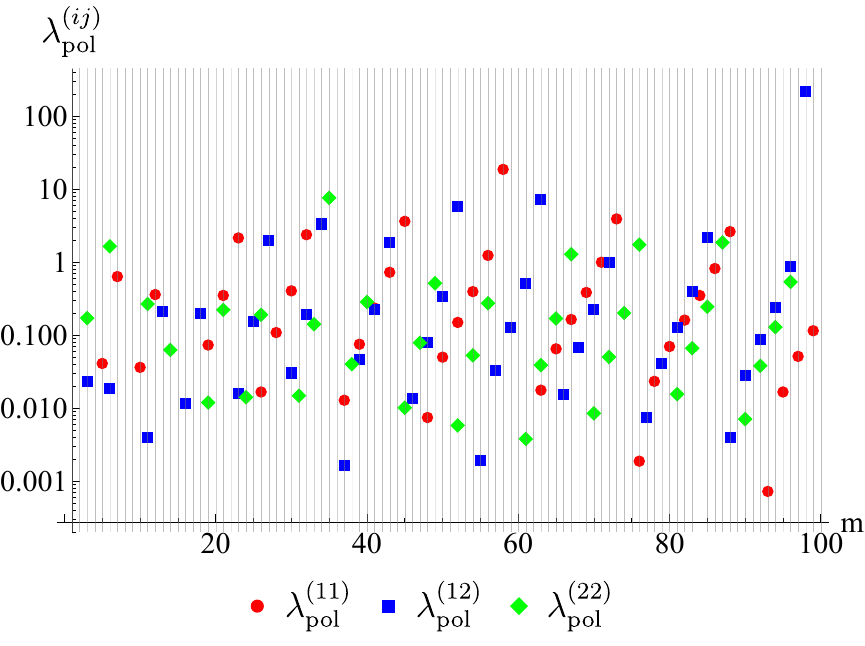}}
  \captionsetup{font=small} 
  \caption{Poles $\lambda^{(ij)}_{\textrm{pol}}$ of the Padé approximant $P_m [g_{ij}^{\scriptscriptstyle (0)}]$ for $m=1,\dots,100$. We have fixed $k=1$.}    
  \label{fig:polospade}
\end{figure}

We compute the Padé approximants $P_m [g_{ij}^{\scriptscriptstyle (0)}]$ of the truncated series of each metric component \eqref{metricacuartico} for $m=1,\dots,100$. To determine the real poles of $P_m [g_{ij}^{\scriptscriptstyle (0)}]$, denoted by $\lambda^{(ij)}_{\textrm{pol}}$, we fix $k=1$. As expected, the number of poles depends on $m$; some Padé approximants have no poles, while others display a single real pole. We find that for $m=1,\dots,100$, $P_m [g_{11}^{\scriptscriptstyle (0)}]$ has 39 poles, $P_m [g_{12}^{\scriptscriptstyle (0)}]$ has 40, and $P_m [g_{22}^{\scriptscriptstyle (0)}]$ has 36. This is shown in Fig.~\ref{fig:polospade}. As we can see, unlike the case of the energy, the Padé approximants obtained for the QMT components, $P_m [g_{ij}^{\scriptscriptstyle (0)}]$, do not show an evident asymptotic pattern or recurring poles at fixed positions, at least within the orders considered. This suggests that the analytic structure of the metric differs from that of the energy. Such behavior could be due either to the absence of a nearby dominant singularity or to a more intricate distribution of singularities in the complex $k$-plane, which could be consistent with the fact that the computation of the metric involves derivatives of the state, and it may be sensitive to subtler aspects of the system.

Comparing Padé approximants with and without poles shows that the latter provide more accurate results: whenever a pole is present, the approximation becomes unreliable in its vicinity. 

\begin{figure}[H]
  \centering
  \begin{subfigure}{0.44\linewidth}{\includegraphics[width=1\linewidth]{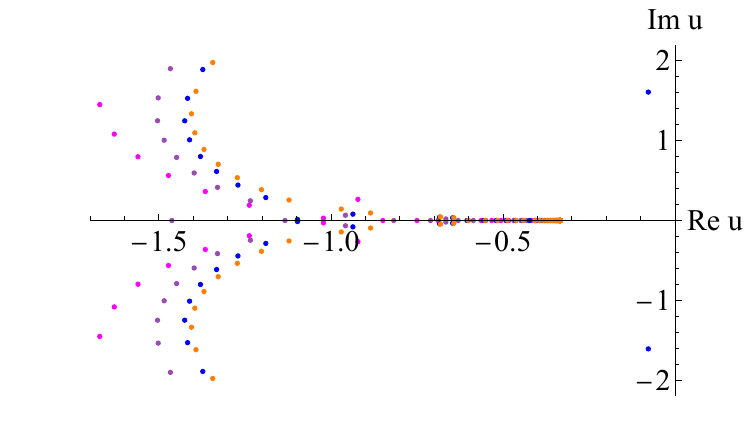}}
    \caption{}
  \end{subfigure}
  \begin{subfigure}{0.55\linewidth}\includegraphics[width=1\linewidth]{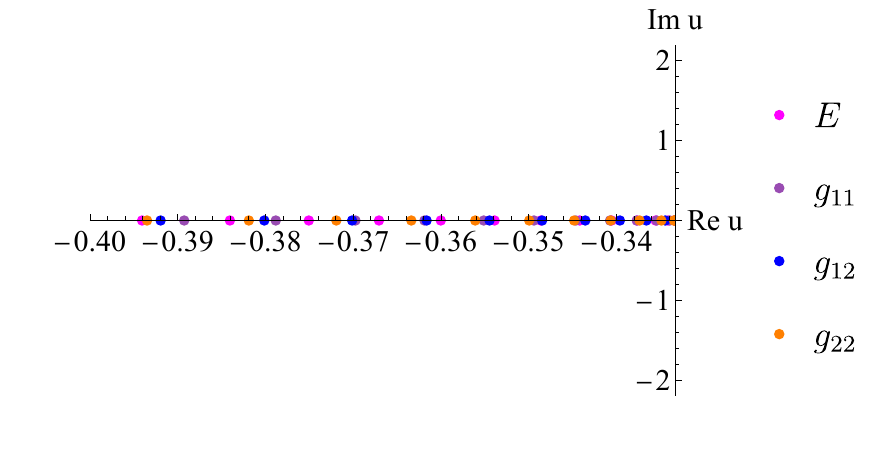}
    \caption{} 
  \end{subfigure}
  \captionsetup{font=small} 
  \caption{(a) Singularities in the Borel plane of $\mathcal{P}_m[E^{(0)}]$ and $\mathcal{P}_m[g_{ij}^{\scriptscriptstyle (0)}]$. An accumulation point is observed around $-1/3$.  (b) Zoom of the figure (a) at the accumulation point.}    
  \label{fig:singularidadesborel}
\end{figure}

In the case of the ground state, we show for $\mathcal{P}_m[E^{(0)}]$ and the three metric components $\mathcal{P}_m[g_{ij}^{\scriptscriptstyle (0)}]$ the distribution of poles (reals and imaginaries) with $k=1$ for $m=100$ in Fig.~\ref{fig:singularidadesborel}.  Notice that both exhibit an accumulation of poles around $-1/3$. This indicates that the instanton contribution to the QMT of the Hamiltonian \eqref{cuartico} with $\lambda<0$ is proportional to $\exp(1/3\lambda)$, consistent with the energy case.

Next, we apply Borel resummation to the Padé approximant of the metric components, i.e.,  $\mathcal{TP}_m[g_{ij}^{{\scriptscriptstyle (0)}}]$. As in the case of the Padé analysis of $g_{ij}$, the Borel–Padé resummation $\mathcal{TP}_m[g_{ij}^{{\scriptscriptstyle (0)}}]$ may or may not display poles depending on the truncation order $m$. In particular, besides negative-real poles consistent with the expected structure (associated with the dominant singularity at $u=-1/3$), spurious poles with $\mathrm{Re}\,u>0$ appear. These are artifacts of the rational approximation (so-called \emph{Froissart doublets}), often accompanied by nearby zeros that cancel their effect \cite{beckermann2016,gilewicz2003}. 

For alternating series and integration contours along $[0,\infty)$, the exact Borel transform has no singularities on the positive axis \cite{SeroneSpadaVilladoro2017}; hence, the poles produced by Padé are spurious and should not be included as physical contributions. Indeed, artificially retaining such residues spoils the accuracy. Consequently, for alternating series such as the present case, Borel resummation must be performed with the principal value (PV) prescription of the Laplace integral, 
\begin{align}
\mathcal{TP}_m\left[g_{ij}^{\scriptscriptstyle (N)}\right] (\lambda)
&= \frac{1}{\lambda}\,\mathrm{PV}\!\int_{0}^{\infty} \mathcal{P}_m\!\left[g_{ij}^{\scriptscriptstyle (N)}\right](u)\,e^{-u/\lambda}\,\mathrm{d}u \nonumber \\
&= \lim_{\epsilon\to 0^+}\frac{1}{\lambda}\sum_{r=1}^{\ell}\int_{u_r+\epsilon}^{u_{r+1}-\epsilon}
\mathcal{P}_m\!\left[g_{ij}^{\scriptscriptstyle (N)}\right](u)\,e^{-u/\lambda}\, \mathrm{d}u\,, \label{PVmetricres}
\end{align}
where the intervals $(u_r,u_{r+1})$ exclude only the possible singularities lying on the integration axis. Complex poles depending on $m$ are interpreted as spurious and disregarded. 

\begin{figure}[H]
  \centering
  \begin{subfigure}{0.49\linewidth}{\includegraphics[width=1\linewidth]{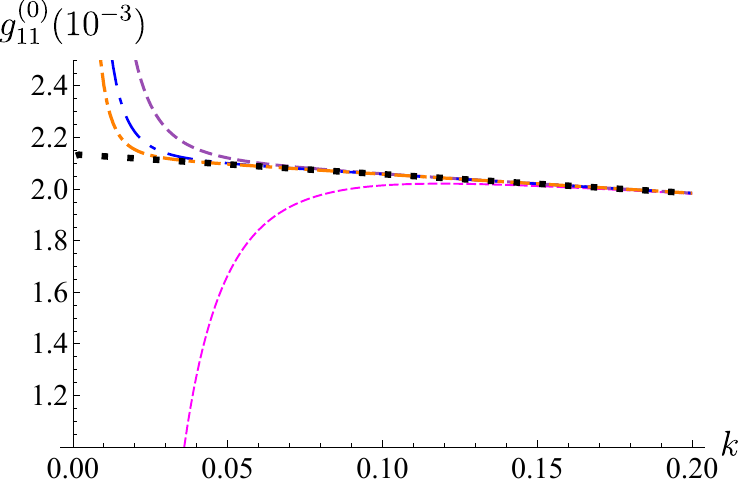}}
    \caption{}
  \end{subfigure}
  \begin{subfigure}{0.49\linewidth}\includegraphics[width=1\linewidth]{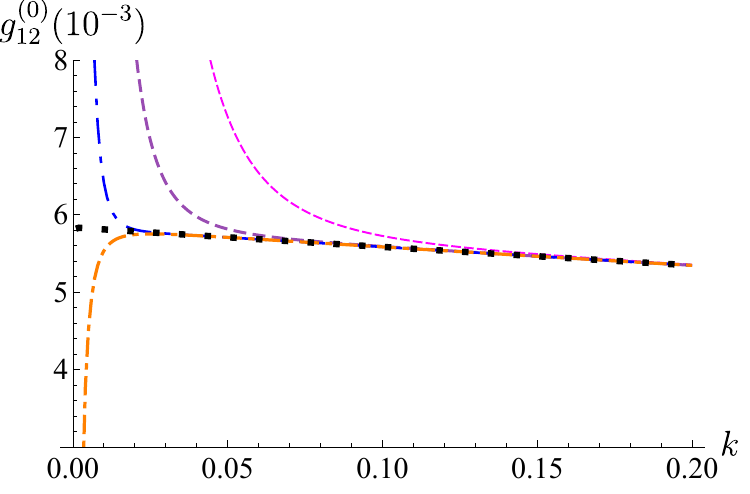}
    \caption{} 
  \end{subfigure}
   \begin{subfigure}{0.49\linewidth}\includegraphics[width=1\linewidth]{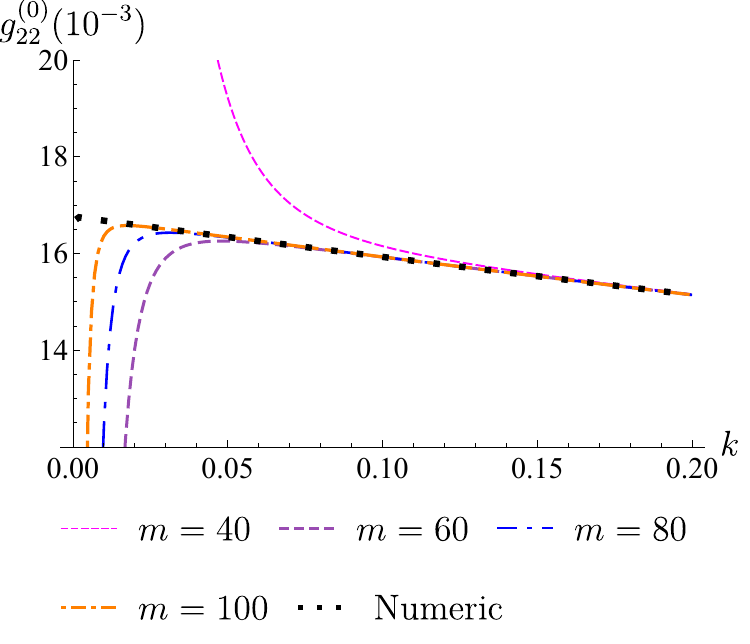}
    \caption{}
  \end{subfigure}
  \captionsetup{font=small} 
  \caption{Metric components obtained by Borel resummation $\mathcal{TP}_m\left[g_{ij}^{\scriptscriptstyle (0)}(\lambda)\right]$ for different truncation orders $m$. As the figures show, the approximation improves as $m$ increases. (a) $g_{11}^{\scriptscriptstyle (0)}$ component, (b) $g_{12}^{\scriptscriptstyle (0)}$ component, and (c) $g_{22}^{\scriptscriptstyle (0)}$ component.}    
  \label{fig:gres}
\end{figure}

We then compare the metric components obtained by Borel resummation \eqref{PVmetricres}  with the exact result from Hamiltonian diagonalization. Fixing $\lambda=1$ and varying $k\in[0.001,0.2]$ in steps of $0.001$, we find that Borel resummation $\mathcal{TP}_m\left[g_{ij}^{\scriptscriptstyle (0)}(\lambda)\right]$ converges more closely to the exact value as more terms $m$ are included, although discrepancies increase as $k\to 0$. These results are shown in Fig. \ref{fig:gres}. 

\subsubsection{Excited states}

The perturbative series depends on the quantum number $N$, making the study of excited states necessary and relevant. In particular, the matrix elements of the operators $\hat{\mathcal{O}}_1 = \tfrac{1}{2}\hat{q}^2$ and $\hat{\mathcal{O}}_2 = \hat{q}^4$ (used to compute the QMT, see \eqref{QMTpert}) grow in magnitude with $N$. As a consequence, the validity range of Borel resummation is expected to shrink for larger $N$.

\begin{figure}[H]
  \centering
  \begin{subfigure}{0.49\linewidth}{\includegraphics[width=1\linewidth]{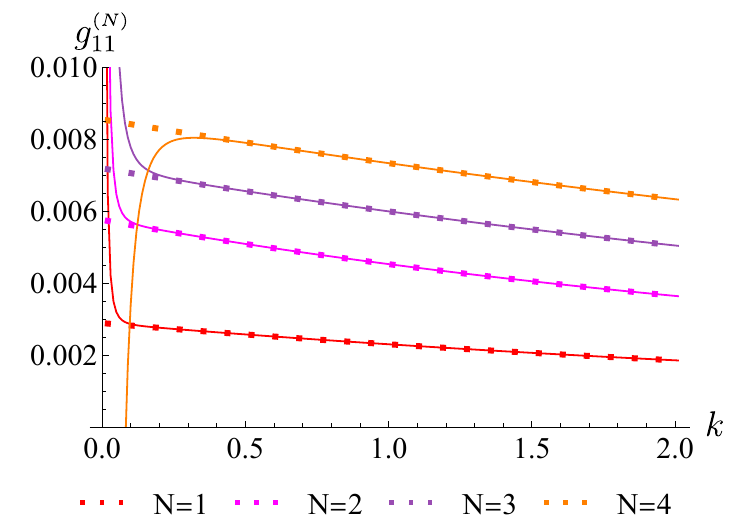}}
    \caption{}
  \end{subfigure}
  \begin{subfigure}{0.49\linewidth}\includegraphics[width=1\linewidth]{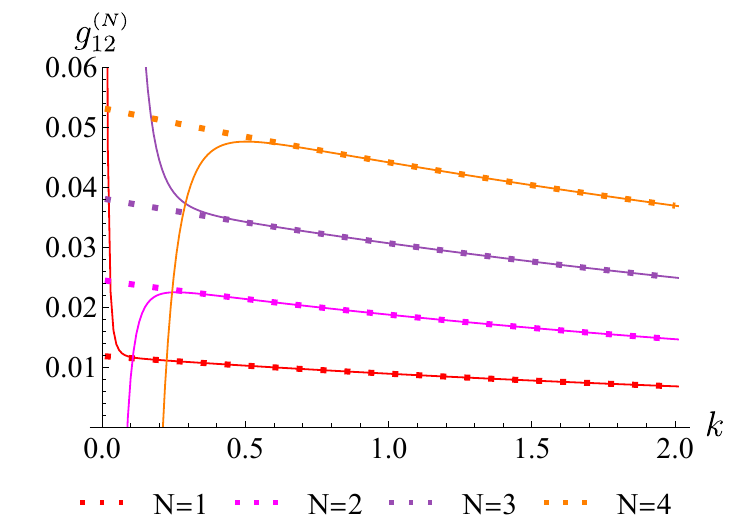}
    \caption{} 
  \end{subfigure}
   \begin{subfigure}{0.49\linewidth}\includegraphics[width=1\linewidth]{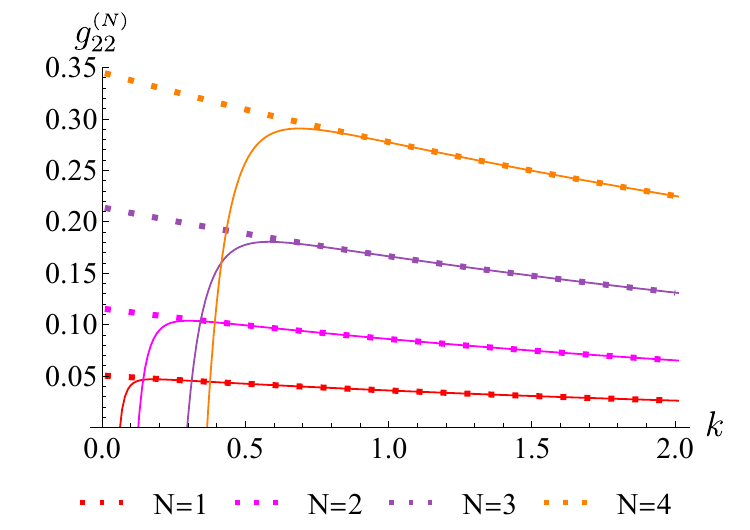}
    \caption{}
  \end{subfigure}
  \captionsetup{font=small} 
  \caption{Metric components obtained by Borel resummation $\mathcal{TP}_{m}[g_{ij}^{\scriptscriptstyle (N)}]$ (solid line) and by exact diagonalization (dashed line), for the exited states $N=1,2,3,4$. (a) $g_{11}^{\scriptscriptstyle (N)}$ component, (b) $g_{12}^{\scriptscriptstyle (N)}$ component, and (c) $g_{22}^{\scriptscriptstyle (N)}$ component.} 
  \label{fig:gresexc}
\end{figure}

For computational reasons, we restrict our analyses to the first four excited states, and the corresponding perturbative series were truncated at order $m=50$. The Borel resummation was carried out using the same Padé–Borel principal value prescription \eqref{PVmetricres}, as in the ground state. We compare our results with those obtained by exact calculations using numerical diagonalization of the Hamiltonian \eqref{cuartico} in a harmonic oscillator basis, ensuring convergence of both eigenvalues and eigenvectors. In Fig.~\ref{fig:gresexc}, we show the resummed metric components alongside the exact values. As in the ground state, we observe that as $N$ increases, the accuracy of the resummation deteriorates with respect to the numerical calculation for small $k$. 

A large-order asymptotic analysis confirms that the location of the dominant Borel singularity remains fixed at $u=-1/3$ for all excited states considered, so the non-perturbative scale is the same as in the ground state. However, the exponents $\beta$ governing the factorial growth of the coefficients increase with $N$, producing proportionally larger non-perturbative contributions. This naturally explains why the validity range of Borel resummation shrinks as the quantum number $N$ grows.

Having analyzed the quartic oscillator in detail, both at the level of the energy spectrum and the quantum metric tensor, we now turn to a higher-order anharmonic system. The next example is the sextic oscillator, which further illustrates the interplay between asymptotics, Borel resummation, and resurgence.

\section{Sextic anharmonic oscillator}\label{sec:sextic}

As a second illustrative example, we consider the sextic potential, whose Hamiltonian is given by
\begin{align}
\label{sextic}
    \hat{H}=\frac{\hat{p}^2}{2}+\frac{k \hat{q}^2}{2}+\lambda \hat{q}^6\,,
\end{align}
with parameters $(k,\lambda)$ defined analogously to the quartic case. The sextic oscillator has been widely studied as a prototypical higher-order anharmonic system. Its perturbative expansion of the energy defines a Borel--Leroy series of order $2$, i.e., the asymptotic growth of the ground-state coefficients is known to be~\cite{Jentschura:2010zza}
\begin{align}
\label{asymEn}
    a^{(0)}_n \sim (-1)^{n+1}\frac{2^{5n+5/2}}{\pi^{2n+2}}\Gamma(2n+1/2)\,.
\end{align}
The factorial divergence is thus stronger than in the quartic case, reflecting the richer large-order structure of the problem.  For this series, if we want to use resurgence, we need to apply the generalized ($\alpha$-)Borel transform (see \eqref{eq:alphaBorel} and \eqref{eq:alphaInverse}).

For the computation of the QMT, we use the \texttt{BenderWu} package again. We calculate the wave function up to order 100 in $\lambda$ and obtain the QMT to the same order. The components of the QMT take the form
\begin{subequations} 
\label{metricasexto}
\begin{align}
    g_{11}^{\scriptscriptstyle (N)}&=\frac{1}{k^2}\sum_{n=0}^{\infty}c^{\scriptscriptstyle (N;n)}_{11}\left(- \frac{\lambda}{k^{2}}\right)^{n} \, ,\\
    g_{12}^{\scriptscriptstyle (N)}&=\frac{1}{k^{5/2}}\sum_{n=0}^{\infty}c^{\scriptscriptstyle (N;n)}_{12}\left(- \frac{\lambda}{k^{2}}\right)^{n} \,,\\
    g_{22}^{\scriptscriptstyle (N)}&=\frac{1}{k^{3}}\sum_{n=0}^{\infty}c^{\scriptscriptstyle (N;n)}_{22}\left(- \frac{\lambda}{k^{2}}\right)^{n}\,.
\end{align}
\end{subequations}
The first coefficients for the energy $a^{(n)}$ and the metric $c_{ij}^{(n)}$ are explicitly shown in Appendix ~\ref{coefsex}.
As in the case of energy, the QMT coefficients lead to a Borel--Leroy series of order 2. Their large-order asymptotics are given by
\begin{subequations}
\label{coef:sexto}
    \begin{align}
    c_{11}^{(n)}& \sim  0.24432537250950578119\,\frac{2^{5n}}{\pi^{2n}}\Gamma(2n+5/2)\,,\\
     c_{12}^{(n)}& \sim 0.30194070537834116441\,\frac{2^{5n}}{\pi^{2n}}\Gamma(2n+9/2)\,,\\
     c_{22}^{(n)}& \sim 0.28599238156715138322\,\frac{2^{5n}}{\pi^{2n}}\Gamma(2n+13/2)\,,   
\end{align}
\end{subequations}
with numerical prefactors determined to 20-digit precision. Unlike the quartic case~\eqref{coef:cuartico}, here we could not express the prefactors as simple rational numbers.

For the numerical implementation of the Borel--Leroy resummation, we considered the transform \eqref{eq:alphaBorel} with $\alpha=2$ and a variable $\beta$. Then, instead of imposing $\beta$ directly from the large-order growth of the perturbative coefficients [cf.~Eqs.~\eqref{asymEn}, \eqref{coef:sexto}], we adopt a variational numerical criterion:
\begin{enumerate}
    \item At a parameter value where the exact (or highly accurate numerical) value of a physical observable $f$ (energy or QMT component) is known, compute the resummation $\mathcal{TP}^{(2,\beta)}_m[f]$ for different $\beta$.
    \item Determine the optimal $\beta^*$ that minimizes the absolute error
    \begin{equation}
        \Delta(\beta) \;=\; \left| f_{\text{exact}} - \mathcal{TP}^{(2,\beta)}_m[f] \right|\,.
    \end{equation}
    \item Use this $\beta^*$ not only at that point but across the entire parameter domain. 
\end{enumerate}

\begin{figure}[H]
  \centering
  \begin{subfigure}{0.49\linewidth}{\includegraphics[width=1\linewidth]{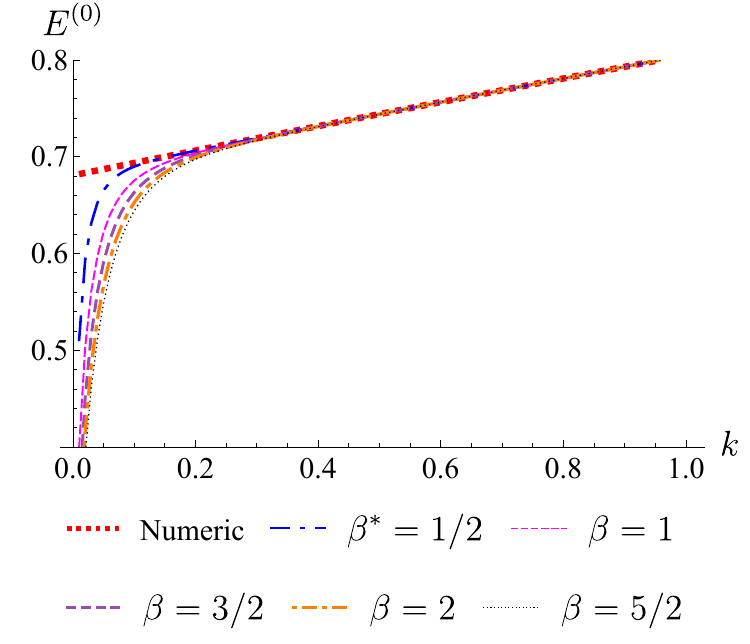}}
    \caption{}
  \end{subfigure}
  \begin{subfigure}{0.49\linewidth}{\includegraphics[width=1\linewidth]{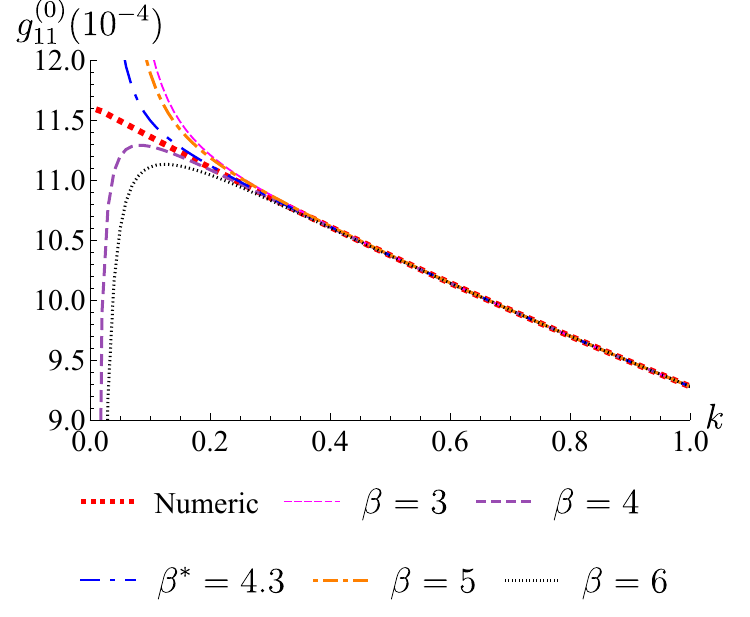}}
    \caption{}
  \end{subfigure}
  \begin{subfigure}{0.49\linewidth}\includegraphics[width=1\linewidth]{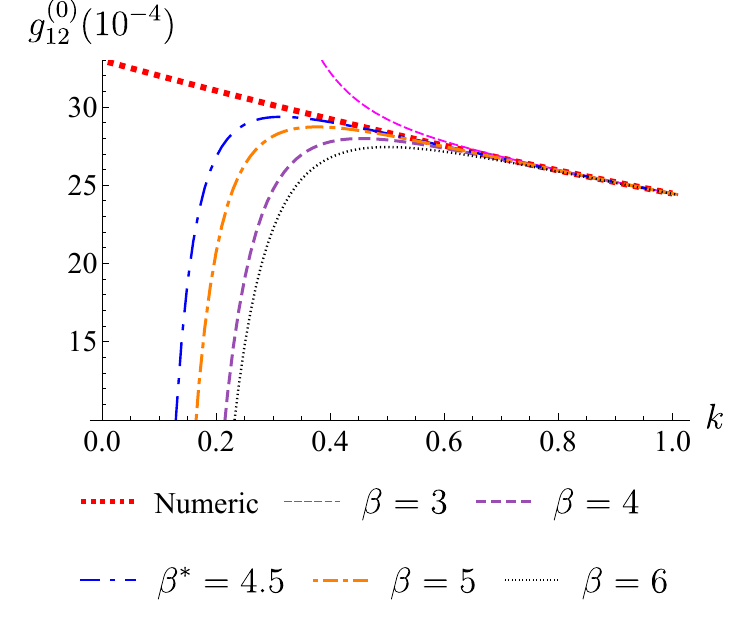}
    \caption{} 
  \end{subfigure}
   \begin{subfigure}{0.49\linewidth}\includegraphics[width=1\linewidth]{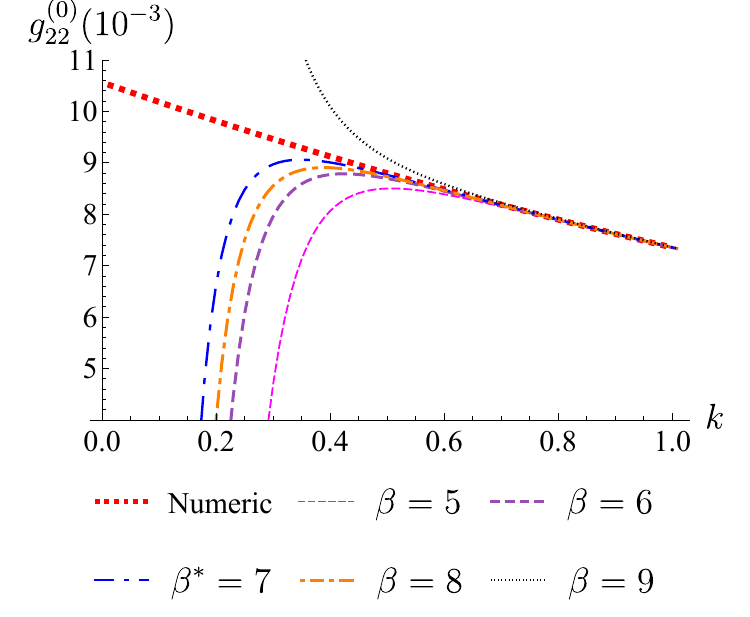}
    \caption{}
  \end{subfigure}
  \captionsetup{font=small} 
  \caption{Energy and QMT components obtained by Borel--Leroy resummation for the ground state, $\mathcal{TP}_m[E^{(0)}]$ and  $\mathcal{TP}_m[g_{ij}^{(0)}]$, for different values of $\beta$. Notice that the one corresponding to the respective $\beta^*$ is the best approximation. (a) $E^{(0)}$, (b) $g_{11}^{\scriptscriptstyle (0)}$ component, (c) $g_{12}^{\scriptscriptstyle (0)}$ component, and (d) $g_{22}^{\scriptscriptstyle (0)}$ component.}    
  \label{fig:gres6}
\end{figure}

We compare the corresponding resummation obtained using $\beta^*$ with other choices of $\beta$. In our analysis we fixed $\lambda=1$ and varied $k\in(0,1)$. The results for the energy eigenvalues and QMT coefficients are shown in Fig.~\ref{fig:gres6}, where the optimal $\beta^*$ is plotted as a blue dashed line.

The optimal value $\beta^*$ is not identical to the one predicted by the large-order asymptotics of the perturbative coefficients. Indeed, the asymptotic analysis yields
\begin{equation}
a_n \sim \mathcal{A}\, S^{-n}\,\Gamma(2n+\beta)\,\Big(1+\frac{c_1}{n}+\cdots\Big)\,,
\end{equation}
suggesting that the ``natural'' Borel--Leroy parameter is $\beta^*=\beta$. However, in practice, when working with a finite number of coefficients and Padé--Borel resummations affected by spurious poles and $1/n$ corrections, the numerically optimal value that minimizes the error deviates from $\beta$, i.e., it acts as an effective parameter compensating for finite-order corrections and numerical details of the resummation. The Fig. \ref{fig:gres6} shows that the proposed numerical implementations improve the results.

The sextic oscillator thus illustrates how generalized Borel--Leroy techniques naturally arise once the large-order growth of perturbative coefficients deviates from the standard $n!$ behavior. In particular, the presence of $\Gamma(2n+\beta)$ growth requires the use of $\alpha=2$-Borel transforms, and the associated resummations provide quantitatively accurate results even in the presence of strong anharmonicities. Having established the applicability of resurgence methods beyond the quartic case, we now turn to a different generalization: anharmonic oscillators in higher spatial dimensions. The $d$-dimensional quartic oscillator with spherical symmetry serves as a natural prototype for exploring how the analytic structure of perturbative expansions and their Borel resummation extend to systems with richer configuration spaces.

\section{The $d$-dimensional quartic anharmonic oscillator}\label{sec:ddim}

The techniques of resummation of divergent series have proven to be fundamental and useful tools in the analysis of quantum systems, whose perturbative expansion is asymptotic rather than convergent. We have analyzed one-dimensional systems; extending our study to higher-dimensional settings opens new possibilities for the non-perturbative study of physically relevant models.

As a concrete example of this generalization, we consider the $d$-dimensional quartic oscillator with spherical symmetry, whose Hamiltonian reads
\begin{equation}
\label{cuartico3d}
\hat{H} = \frac{\hat{p}^2}{2} + \frac{k \hat{r}^2}{2} + \lambda \hat{r}^4\,,
\end{equation}
where $\hat{r} = || \hat{\boldsymbol{r}}||$ and $\hat{\boldsymbol{r}}=(\hat{x}_1,\dots,\hat{x}_d)$ is the position operator in $d$ dimensions. This system describes an isotropic harmonic oscillator subject to a quartic perturbation, and serves as a prototype to analyze the singularity structure and non-perturbative effects in higher-dimensional configuration spaces.

Because of the spherical symmetry of the potential, the problem reduces to an effective radial equation that depends on the angular momentum. The construction of perturbative expansions for the $d$-dimensional quartic oscillator and the analysis of their large-order growth have been studied in detail in~\cite{delValle2019}. They also show that the perturbative expansion of the energy levels in powers of $\lambda$ is divergent but Borel summable, allowing for its study. We must mention that variational perturbation theory has also been applied to this model to obtain high-order perturbative series and study their analytic properties~\cite{Brandt}.

For perturbative analysis, we separate the Hamiltonian into a solvable part $\hat{H}_0$ and an interaction term $\hat{H}_1$, with
\begin{subequations}
\begin{align}
    \hat{H}_0&=\frac{\hat{p}^2}{2}+\frac{k \hat{r}^2}{2}\,,\\
    \hat{H}_1&=\lambda \hat{r}^4\,.
\end{align}
\end{subequations}
The Schrödinger equation for $H_0$ is well known \cite{messiah1999}. Writing the Laplacian in hyperspherical coordinates, one obtains
\begin{align}
    -\frac{1}{2}\left( \nabla^2+\frac{k r^2}{2}  \right)\Psi(r,\theta)=E\Psi(r,\theta)\,,
\end{align}
with $\theta=(\theta_1,\dots,\theta_{d-1})$. For radial wave functions with angular quantum number $l=0$, the radial part $R(r)$ satisfies
\begin{align}
   - \frac{1}{2r^{d-1}} \frac{d}{dr} \left( r^{d-1} \frac{dR (r)}{dr} \right)  + \frac{k}{2} r^2 R(r) = E R(r)\,,
\end{align}
whose normalized solutions are given by
\begin{align}
   R_{N}(r) =
\sqrt{ \frac{2 \, \omega^{ \frac{d}{2}} \, N! }{ \Gamma(N + \frac{d}{2}) } } \exp\left({ -\frac{\omega}{2} r^2 }\right) \,
L_N^{\frac{d}{2}-1}( \omega r^2 )\,,
\end{align}
where $L_{N}^{d/2-1}$ are the associated Laguerre polynomials.

Starting from the exact solutions of the free oscillator, we treat the quartic term perturbatively in $\lambda$. Since the interaction depends only on the radial coordinate, the perturbation is diagonal in the spherical harmonic basis, simplifying the computation of energy corrections and wave functions. 

\begin{figure}[H]
  \centering
  \begin{subfigure}{0.49\linewidth}{\includegraphics[width=1\linewidth]{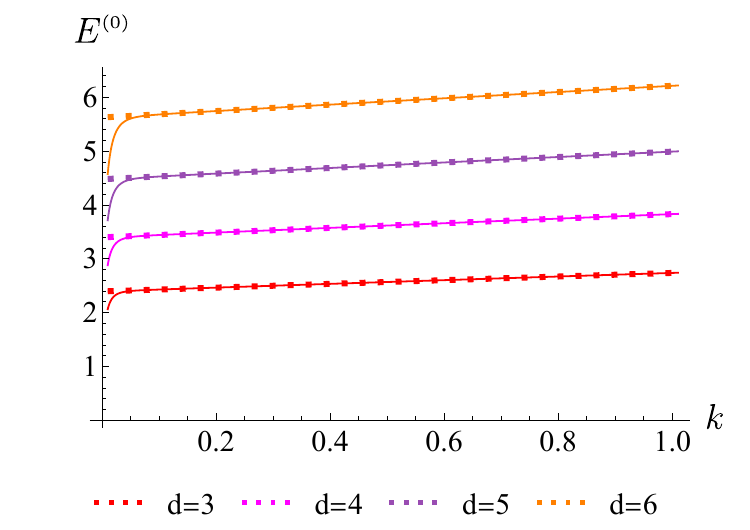}}
    \caption{}
  \end{subfigure}
  \begin{subfigure}{0.49\linewidth}{\includegraphics[width=1\linewidth]{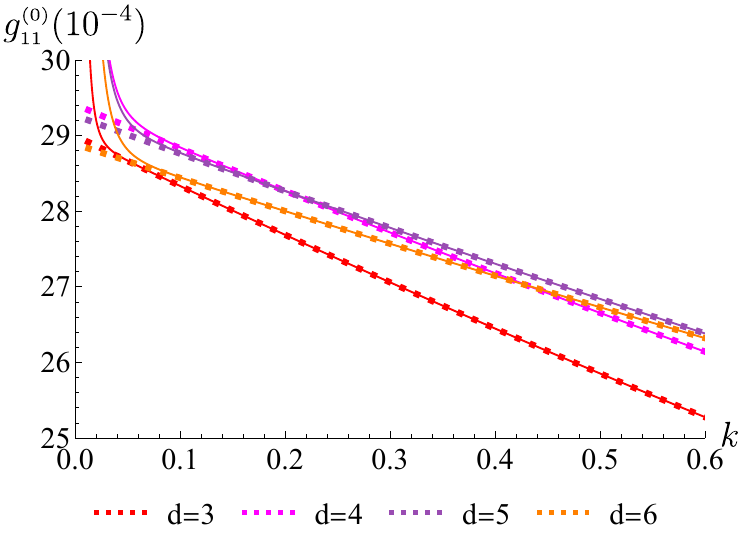}}
    \caption{}
  \end{subfigure}
  \begin{subfigure}{0.49\linewidth}\includegraphics[width=1\linewidth]{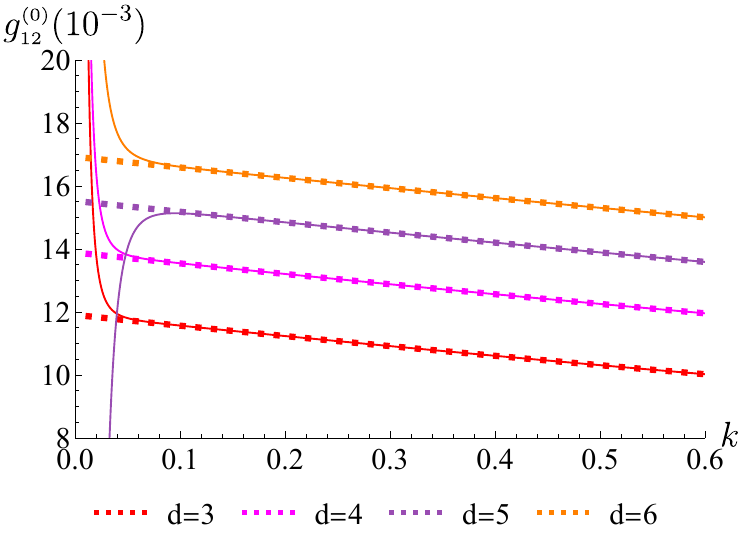}
    \caption{} 
  \end{subfigure}
   \begin{subfigure}{0.49\linewidth}\includegraphics[width=1\linewidth]{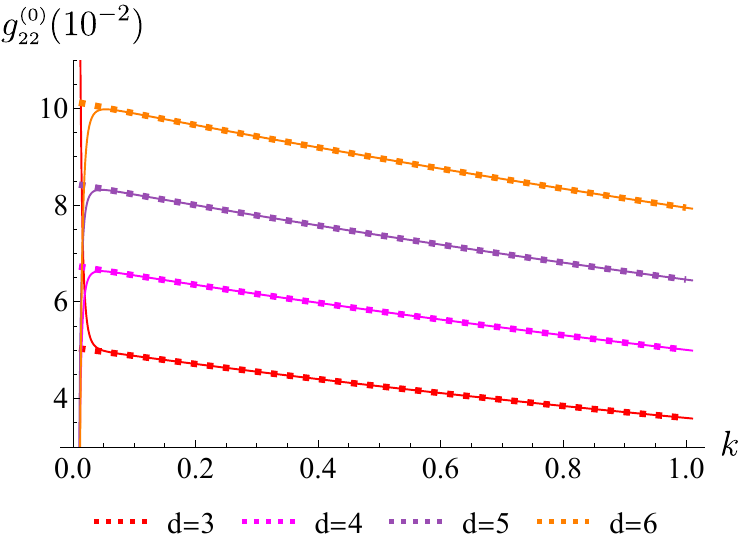}
    \caption{}
  \end{subfigure}
  \captionsetup{font=small} 
  \caption{Comparison between the energy and QMT components obtained via Borel resummation,$\mathcal{TP}_m\left[E^{(0;d)}\right]$ and  $\mathcal{TP}_m\left[g_{ij}^{(0;d)}\right]$, respectively, for $m=100$ (solid line) and numerical results obtained with the Lagrange Mesh method (dashed line) for $d=3,4,5,6$. (a) Energy $E^{(0)}$, (b) $g_{11}^{\scriptscriptstyle (0)}$ component, (c) $g_{12}^{\scriptscriptstyle (0)}$ component, and (d) $g_{22}^{\scriptscriptstyle (0)}$ component.}
  \label{fig:gres3d}
\end{figure}

Using Provost’s formula \eqref{QMTpert}, the components of the QMT take the same functional form as in Eq.~\eqref{metricacuartico}. For the ground state, the first coefficients of the energy $a^{(n)}$ and the metric $c_{ij}^{(n)}:=c_{ij}^{(0,n)}$ are collected in Appendix~\ref{coefddim}, and to distinguish between different spatial dimensions, we introduce the notation  $E^{(0; d)}$ and $g_{ij}^{\scriptscriptstyle(0;d)}$, where $d$ denotes the dimension, for the energy and for the QMT, respectively. For numerical computations of both energy and metric, we employed the \texttt{LagrangeMesh} package~\cite{delValle2022}, using a mesh of 200 points. In Fig.~\ref{fig:gres3d} we compare the corresponding numerical results (dashed lines) with Borel resummations $\mathcal{TP}_m\left[E^{(0; d)}\right]$ and $\mathcal{TP}_m\left[g_{ij}^{\scriptscriptstyle(0;d)}\right]$ obtained with $m=100$ (solid lines). We observe that the Borel resummation reproduces the exact values more accurately as $k$ increases, or equivalently, for smaller $\lambda$.

We find that the large-order growth of the perturbative coefficients is 
\begin{subequations}
\label{coef:cuarticodd}
    \begin{align}
    a^{(n)} &\sim C \cdot 3^{n +\frac{d}{2}}\Gamma\left(n+\frac{d}{2}\right)\,,\\
    c_{11}^{(n)}& \sim C_{11} \cdot 3^{n +\frac{d}{2}} \Gamma \left(n +2+\frac{d}{2}\right)\,,\\
    c_{12}^{(n)}& \sim C_{12} \cdot 3^{n +\frac{d}{2}}\Gamma \left(n+3+\frac{d}{2}\right)\,,\\
    c_{22}^{(n)}& \sim C_{22} \cdot 3^{n +\frac{d}{2}}\Gamma \left(n+4+\frac{d}{2}\right)\,,
\end{align}
\end{subequations}
where $C$ and $C_{ij}$ are dimension-dependent constants independent of $n$. For the energy $E^{(0;d)}$ and the QMT components $g_{12}^{\scriptscriptstyle (0;d)}$ and $g_{22}^{\scriptscriptstyle (0;d)}$, we find that, at fixed values of the parameters $x=(k,\lambda)$, one has
\begin{align}
g_{ij}^{(0;d+1)} > g_{ij}^{(0;d)} \, .
\end{align}
On the other hand, the component $g_{11}^{(0;d)}$ exhibits crossings between different dimensions $d$.

Both the energy and the QMT components reproduce the numerical results with remarkable accuracy. This confirms that Borel resummation is a particularly effective tool for extracting physical information from asymptotic series in higher-dimensional quantum systems.

The $d$-dimensional quartic oscillator thus provides a natural testing ground for the extension of resurgence techniques beyond one-dimensional systems. Both the energy spectrum and the quantum metric tensor exhibit factorially divergent perturbative expansions whose large-order growth encodes non-perturbative scales, in full analogy with the one-dimensional case. The Borel–Padé resummation was shown to reproduce with high accuracy the exact numerical results obtained through independent methods, confirming the robustness of the approach in higher dimensions. At the same time, the dimensional dependence introduces systematic modifications in the asymptotic laws of the perturbative coefficients, highlighting the richer analytic structure of multidimensional systems. These results strengthen the case for resurgence and generalized Borel techniques as versatile tools in quantum mechanics, capable of bridging perturbative and non-perturbative physics across various settings. 

\section{Conclusions}\label{sec:conclusions}

Here, we summarize the main findings of this work and outline possible directions for further research. We have investigated the role of resurgence and Borel resummation techniques in the perturbative analysis of quantum systems, with particular emphasis on the quantum metric tensor. Starting from the general framework of asymptotic series and generalized Borel transforms, we have illustrated how non-perturbative information is encoded in the large-order growth of perturbative coefficients, and how suitable resummation procedures enable the extraction of accurate physical predictions from otherwise divergent expansions.

As a first case study, we considered the quartic oscillator, a paradigmatic system whose perturbative expansions are divergent but asymptotic. By analyzing both the energy levels and the components of the quantum metric tensor, we showed that the factorial growth of the coefficients and the associated singularity structure are precisely the features that make the model a textbook example of resurgence. The comparison between Padé approximants, Borel--Padé resummations, and exact diagonalization confirmed the accuracy and consistency of the method, while also clarifying the role of spurious poles and the importance of the principal value prescription for alternating series.

The sextic oscillator provided a natural extension, exhibiting more intricate large-order behavior characterized by $\Gamma(2n+\beta)$ growth. In this case, the standard Borel transform is not sufficient, and one must employ generalized Borel–Leroy techniques with $\alpha=2$. We have implemented these resummations numerically, showing that they reproduce the exact results with high accuracy and highlighting the flexibility of the formalism in adapting to different asymptotic growth laws. The sextic model, therefore, demonstrates how generalized Borel transforms emerge as the natural tool once perturbative coefficients deviate from the standard Gevrey-1 case.

Finally, we extended the analysis to the $d$-dimensional quartic oscillator with spherical symmetry. This example illustrates the application of resurgence methods to higher-dimensional systems, where both the energy and the quantum metric tensor remain factorially divergent but Borel summable. The dimensional dependence was shown to systematically modify the asymptotic laws of the coefficients, introducing richer analytic structures while preserving the overall resurgence framework. Numerical comparisons with results obtained via the Lagrange Mesh method confirmed that Borel–Padé resummations provide quantitatively accurate predictions also in multidimensional settings.

In all the examples, Borel's resummation of the energy works better than the ones of the QMT components; we consider that this is because the growth of the coefficients of the QMT components series is bigger than that of the energy (cf. for example \eqref{eq:coefenergy} and \eqref{eq:coeffQMT}). We remark that we have not analyzed Berry's curvature because it is zero in the considered examples. We plan to study other systems with non-trivial Berry curvature, extending our work to the QGT.  

Taken together, these results reinforce the idea that resurgence and Borel resummation techniques are powerful and versatile tools in quantum mechanics and quantum geometry. They not only recover exact results from divergent series but also unveil the underlying analytic structure of the problem, bridging perturbative and non-perturbative physics. Beyond the models considered here, our methods can be extended to more general families of anharmonic oscillators, quantum field theoretical models, and systems with explicit time dependence. In this last case, it will be necessary to apply the analysis on the Time-dependent quantum geometric tensor \cite{Diaz2025}. Another interesting direction for future research is the exploration of the interplay between the quantum geometric tensor, non-perturbative effects, and entanglement properties, which may further deepen our understanding of quantum geometry in both finite- and infinite-dimensional systems.

\acknowledgments

This work was partially supported by DGAPA-PAPIIT Grant No. IN114225. B. D. acknowledges support from SECIHTI (M\'exico)  No 371778. M. J. H. acknowledges support from SECIHTI (M\'exico)  No 940148.

\appendix
\section{Coefficient of the energy and the quantum metric components }

\subsection{Sextic anharmonic oscillator}\label{coefsex}
\begin{table}[H]
\centering
\setlength{\tabcolsep}{6pt}
\renewcommand{\arraystretch}{1.2}
\begin{tabular}{|l|l l |}
\hline
$n$ & $a^{(n)}$ & $c_{11}^{(n)}$  \\
\hline
0  & $\tfrac{1}{2}$ & $\tfrac{1}{32}$  \\
1  & $\tfrac{15}{8}$ & $\tfrac{315}{128}$ \\
2  & $\tfrac{3495}{64}$ & $\tfrac{347345}{1024}$  \\
3  & $\tfrac{1239675}{256}$ & $\tfrac{294417175}{4096}$ \\
4  & $\tfrac{3342323355}{4096}$ & $\tfrac{1446615522213}{65536}$ \\
5  & $\tfrac{3625381915125}{16384}$ & $\tfrac{2475351224849465}{262144}$ \\
6  & $\tfrac{11569592855303595}{131072}$ & $\tfrac{11410628404292094533}{2097152}$ \\
7  & $\tfrac{25582203502337850075}{524288}$ & $\tfrac{34361543346089807333895}{8388608}$\\
8  & $\tfrac{600122673764873281048275}{16777216}$ & $\tfrac{1052535281392938021184146029}{268435456}$\\
9  & $\tfrac{2255790551335990113153656625}{67108864}$ & $\tfrac{5004773302576178054117081385925}{1073741824}$ \\
10 & $\tfrac{21160722559334931139552067094465}{536870912}$ & $\tfrac{57931578499307990113387472415826671}{8589934592}$  \\
\hline
\end{tabular}
\end{table}

\begin{table}[H]
\centering
\setlength{\tabcolsep}{6pt}
\renewcommand{\arraystretch}{1.2}
\begin{tabular}{|l|l  l|}
\hline
$n$ & $c_{12}^{(n)}$ & $c_{22}^{(n)}$ \\
\hline
0  &  $\tfrac{45}{64}$ & $\tfrac{685}{32}$ \\
1  & $\tfrac{29455}{256}$ & $\tfrac{3179975}{512}$ \\
2  & $\tfrac{52852675}{2048}$ & $\tfrac{8740862727}{4096}$ \\
3  & $\tfrac{66513619563}{8192}$ & $\tfrac{15710791131635}{16384}$ \\
4  & $\tfrac{460478023025435}{131072}$ & $\tfrac{148369375680563899}{262144}$ \\
5  & $\tfrac{1067538422387182081}{524288}$ & $\tfrac{452699174219715674985}{1048576}$ \\
6  & $\tfrac{6450716965087634727255}{4194304}$ & $\tfrac{3494839383194765434541143}{8388608}$ \\
7  &  $\tfrac{24750848222561711978886547}{16777216}$ & $\tfrac{16710616434338738597083363475}{33554432}$ \\
8  & $\tfrac{942903034069071890446545100075}{536870912}$ & $\tfrac{776946886496528372877515770259163}{1073741824}$ \\
9  & $\tfrac{5463285568274353187219758033055877}{2147483648}$ & $\tfrac{5398632325185087894762769126705990205}{4294967296}$ \\
10 & $\tfrac{75741696378287915054620727743303669165}{17179869184}$ & $\tfrac{88435318840083000267181804137171175358285}{34359738368}$ \\
\hline
\end{tabular}
\end{table}
\subsection{Quartic $d$-dimensional anharmonic oscillator} \label{coefddim}
\subsubsection{$d=3$}
\normalsize
\begin{table}[H]
\centering
\setlength{\tabcolsep}{6pt}
\renewcommand{\arraystretch}{1.2}
\begin{tabular}{|l|l l l l|}
\hline
$n$ & $a^{(n)}$ & $c_{11}^{(n)}$ & $c_{12}^{(n)}$ & $c_{22}^{(n)}$ \\
\hline
0  & $\tfrac{3}{2}$ & $\tfrac{3}{32}$ & $\tfrac{15}{16}$ & $\tfrac{315}{32}$ \\
1  & $\tfrac{15}{4}$ & $\tfrac{165}{64}$ & $\tfrac{2115}{64}$ & $\tfrac{855}{2}$ \\
2  & $\tfrac{165}{8}$ & $\tfrac{18585}{256}$ & $\tfrac{137955}{128}$ & $\tfrac{1020485}{64}$ \\
3  & $\tfrac{3915}{16}$ & $\tfrac{1110795}{512}$ & $\tfrac{9167025}{256}$ & $\tfrac{149968485}{256}$ \\
4  & $\tfrac{520485}{128}$ & $\tfrac{281939715}{4096}$ & $\tfrac{2533049415}{2048}$ & $\tfrac{44989504075}{2048}$ \\
5  & $\tfrac{21304485}{256}$ & $\tfrac{18939326085}{8192}$ & $\tfrac{91497398865}{2048}$ & $\tfrac{872413570935}{1024}$ \\
6  & $\tfrac{2026946145}{1024}$ & $\tfrac{2686879002735}{32768}$ & $\tfrac{27707249431995}{16384}$ & $\tfrac{140815286731995}{4096}$ \\
7  & $\tfrac{108603230895}{2048}$ & $\tfrac{200891008804755}{65536}$ & $\tfrac{2199913142248215}{32768}$ & $\tfrac{47446531470324135}{32768}$ \\
8  & $\tfrac{51448922163885}{32768}$ & $\tfrac{126502295750153955}{1048576}$ & $\tfrac{1466115957601162275}{524288}$ & $\tfrac{33433804540644980915}{524288}$ \\
9  & $\tfrac{3325989183831585}{65536}$ & $\tfrac{10474700079759844065}{2097152}$ & $\tfrac{64086302793235712295}{524288}$ & $\tfrac{1541453003571749948775}{524288}$ \\
10 & $\tfrac{465491656557283395}{262144}$ & $\tfrac{1823807507161951314045}{8388608}$ & $\tfrac{23520266222995005656085}{4194304}$ & $\tfrac{297781324592455095279665}{2097152}$ \\
\hline
\end{tabular}
\end{table}

\subsubsection{$d=4$}
\begin{table}[H]
\centering
\setlength{\tabcolsep}{6pt}
\renewcommand{\arraystretch}{1.2}
\begin{tabular}{|l|l l l l|}
\hline
$n$ & $a^{(n)}$ & $c_{11}^{(n)}$ & $c_{12}^{(n)}$ & $c_{22}^{(n)}$ \\
\hline
0  & $2$ & $\tfrac{1}{8}$ & $\tfrac{3}{2}$ & $\tfrac{75}{4}$ \\
1  & $6$ & $\tfrac{33}{8}$ & $\tfrac{501}{8}$ & $954$ \\
2  & $39$ & $\tfrac{2199}{16}$ & $\tfrac{9561}{4}$ & $\tfrac{82487}{2}$ \\
3  & $540$ & $\tfrac{76839}{16}$ & $\tfrac{367803}{4}$ & $\tfrac{3477771}{2}$ \\
4  & $\tfrac{41433}{4}$ & $\tfrac{1410195}{8}$ & $\tfrac{58252869}{16}$ & $\tfrac{296473597}{4}$ \\
5  & $242208$ & $\tfrac{434035107}{64}$ & $\tfrac{1194594951}{8}$ & $3237189093$ \\
6  & $\tfrac{52149999}{8}$ & $\tfrac{4366462011}{16}$ & $\tfrac{203476482105}{32}$ & $\tfrac{291553160727}{2}$ \\
7  & $195776190$ & $\tfrac{1468000954083}{128}$ & $\tfrac{1125661001259}{4}$ & $\tfrac{108657132025821}{16}$ \\
8  & $\tfrac{412171252725}{64}$ & $\tfrac{8045561953443}{16}$ & $\tfrac{3315781923256785}{256}$ & $\tfrac{5247343878773063}{16}$ \\
9  & $229284886527$ & $\tfrac{23541076912140771}{1024}$ & $\tfrac{19847901671834097}{32}$ & $\tfrac{1052280978138682569}{64}$ \\
10 & $\tfrac{1121697677785665}{128}$ & $\tfrac{70121786676976797}{64}$ & $\tfrac{15829040492602273215}{512}$ & $\tfrac{3426487812865879943}{4}$ \\
\hline
\end{tabular}
\end{table}

\subsubsection{$d=5$}
\begin{table}[H]
\centering
\setlength{\tabcolsep}{6pt}
\renewcommand{\arraystretch}{1.2}
\begin{tabular}{|l|l l l l|}
\hline
$n$ & $a^{(n)}$ & $c_{11}^{(n)}$ & $c_{12}^{(n)}$ & $c_{22}^{(n)}$ \\
\hline
0  & $\tfrac{5}{2}$ & $\tfrac{5}{32}$ & $\tfrac{35}{16}$ & $\tfrac{1015}{32}$ \\
1  & $\tfrac{35}{4}$ & $\tfrac{385}{64}$ & $\tfrac{6755}{64}$ & $\tfrac{14805}{8}$ \\
2  & $\tfrac{525}{8}$ & $\tfrac{59255}{256}$ & $\tfrac{590835}{128}$ & $\tfrac{5826975}{64}$ \\
3  & $\tfrac{16625}{16}$ & $\tfrac{4741835}{512}$ & $\tfrac{51690275}{256}$ & $\tfrac{1110379725}{256}$ \\
4  & $\tfrac{2894325}{128}$ & $\tfrac{1582351225}{4096}$ & $\tfrac{18486423375}{2048}$ & $\tfrac{424922917125}{2048}$ \\
5  & $\tfrac{152440575}{256}$ & $\tfrac{137387788125}{8192}$ & $\tfrac{850221880725}{2048}$ & $\tfrac{10344735366525}{1024}$ \\
6  & $\tfrac{18353729625}{1024}$ & $\tfrac{24783991477525}{32768}$ & $\tfrac{322625752557675}{16384}$ & $\tfrac{2063611125605925}{4096}$ \\
7  & $\tfrac{1224596281125}{2048}$ & $\tfrac{2319031611102675}{65536}$ & $\tfrac{31599736487815125}{32768}$ & $\tfrac{846138293865624375}{32768}$ \\
8  & $\tfrac{711224582914125}{32768}$ & $\tfrac{1799190771698567625}{1048576}$ & $\tfrac{25581028488126622875}{524288}$ & $\tfrac{714615202723319669625}{524288}$ \\
9  & $\tfrac{55517570883495875}{65536}$ & $\tfrac{180750329516181511625}{2097152}$ & $\tfrac{1337878928062065384625}{524288}$ & $\tfrac{38905890609035167806375}{524288}$ \\
10 & $\tfrac{9245013891201802875}{262144}$ & $\tfrac{37613298411579926375375}{8388608}$ & $\tfrac{578899755709951781461125}{4194304}$ & $\tfrac{8748866967644068156255875}{2097152}$ \\
\hline
\end{tabular}
\end{table}

\subsubsection{$d=6$}
\begin{table}[H]
\centering
\setlength{\tabcolsep}{6pt}
\renewcommand{\arraystretch}{1.2}
\begin{tabular}{|l |l l l l|}
\hline
$n$ & $a^{(n)}$ & $c_{11}^{(n)}$ & $c_{12}^{(n)}$ & $c_{22}^{(n)}$ \\
\hline
0  & $3$ & $\tfrac{3}{16}$ & $3$ & $\tfrac{99}{2}$ \\
1  & $12$ & $\tfrac{33}{4}$ & $\tfrac{657}{4}$ & $3258$ \\
2  & $102$ & $360$ & $\tfrac{16197}{2}$ & $179788$ \\
3  & $1818$ & $\tfrac{129855}{8}$ & $\tfrac{794253}{2}$ & $\tfrac{19112931}{2}$ \\
4  & $\tfrac{88545}{2}$ & $\tfrac{24277113}{32}$ & $\tfrac{158373945}{8}$ & $\tfrac{4059790871}{8}$ \\
5  & $\tfrac{2595087}{2}$ & $\tfrac{1174706685}{32}$ & $\tfrac{16162038951}{16}$ & $\tfrac{109175231841}{4}$ \\
6  & $\tfrac{172957281}{4}$ & $\tfrac{117498922695}{64}$ & $\tfrac{846274033893}{16}$ & $\tfrac{11969305643085}{8}$ \\
7  & $\tfrac{6355598589}{4}$ & $\tfrac{6065766292299}{64}$ & $2845317206025$ & $\tfrac{1341951089387577}{16}$ \\
8  & $\tfrac{2022705878757}{32}$ & $\tfrac{2583564084224325}{512}$ & $\tfrac{20141051523138303}{128}$ & $\tfrac{616714225435175035}{128}$ \\
9  & $\tfrac{86088409115175}{32}$ & $\tfrac{141783633910817697}{512}$ & $\tfrac{2291221009591376475}{256}$ & $\tfrac{9089294881553064261}{32}$ \\
10 & $\tfrac{7777562767529055}{64}$ & $\tfrac{16037051384522262357}{1024}$ & $\tfrac{134102738605699665693}{256}$ & $\tfrac{550527748853108874043}{32}$ \\
\hline
\end{tabular}
\end{table}

\bibliographystyle{JHEP}
\bibliography{biblio.bib}

\end{document}